\documentclass[11pt,preprint]{aastex}
\usepackage{graphicx}
\usepackage[fleqn]{amsmath}
\usepackage{color}
\usepackage{threeparttable}
\begin{document}

\title{Z Draconis with two companions in a 2:1 mean-motion resonance}

\author{Jinzhao Yuan\altaffilmark{1}, Hakan Volkan \c{S}enavc{\i}\altaffilmark{2}, Juanjuan Liu\altaffilmark{3}, Selim O. SELAM\altaffilmark{2}, and Damla G\"{U}M\"{U}\c{S}\altaffilmark{2}}

\singlespace

\altaffiltext{1}{Department of Physics, Shanxi Normal University, Linfen
041004, Shanxi, China (yuanjz@sxnu.edu.cn)}

\altaffiltext{2}{Department of Astronomy and Space Sciences, Ankara University, Faculty of Science,
 TR-06100 Tando\u{g}an-Ankara, Turkey}

\altaffiltext{2}{Qiaoli Middle School, 041072 Linfen, Shanxi Province, China}

\begin{abstract}

All available mid-eclipse times of the eclipsing binary Z Draconis are analysed, and three sets of cyclic variations with periods of 20.1, 29.96 and 59.88 yr are found. The low-amplitude variations with the period of 20.1 yr may be attributed to the unavoidable and slight imperfection in the double-Keplerian model, which gives the periods of 29.96 and 59.88 yr. Interestingly, the Z Draconis system is close to a 2:1 mean-motion resonance, or 6:3:2 mean-motion resonance if the period of 20.1 yr is true. We also find that the best solutions tend to give the minimum eccentricities. Based on Kepler's third law, the outermost companion has the minimum mass of $\sim0.77M_{\bigodot}$, whereas the middle companion is an M dwarf star with a mass of $\sim0.40M_{\bigodot}$, suggesting that Z Draconis is a general N-body system.

\end{abstract}

\keywords{binaries: close -- stars: individual: Z Draconis -- methods: numerical.}

The oscillations in the mid-eclipse times of eclipsing binaries are usually interpreted as the light-travel time (LTT) effect or the magnetic activity cycles (Applegate 1992; Yuan \& Qian 2007). In the LTT model, a companion revolves around the eclipsing pair. The line-of-sight distance between the eclipsing pair and the barycentre of the whole system, $d$, varies with a strict period equal to the orbital period of the companion. After divided by the speed of light, c, we obtain the $O-C$ value, $d/c$. Obviously, the multi-periodic variations in the eclipse times of an eclipsing binary provide us important constraints on the orbital characteristics of this multi-companion system, which usually comprises an eclipsing binary and multiple sub-stellar objects or planets. In the magnetic activity mechanism, the gravitational or magnetic force changes as the active component goes through a magnetic activity cycle, producing quasi-periodic variations in the eclipse times (Beuermann et al. 2012).

Z Draconis (BD+$73^{\circ}533$ = HIP 57348, $V_{max} = 10.67$ mag) was first found to be an Algol-type binary (hereafter Z Dra AB) by Ceraski (1903). Due to its high declination and brightness, a large number of photometric data were obtained by small telescopes with alt-azimuthal mountings. The first radial velocity curve for the primary component was obtained by Struve (1947). Based on the radial velocity curve and the $BVRI$ light curves obtained with a 0.25 Schmidt-Cassegrain telescope, Terrell (2006) carried out a photometric-spectroscopic analysis. The solutions indicated that Z Dra is a semi-detached binary with masses of 1.47$M_{\bigodot}$ for the primary, and 0.43$M_{\bigodot}$ for the secondary component. Terrell (2006) also pointed out that the mass of the primary is significantly lower than expected for the A5V star, but consistent with the $B-V$ color of $0.45$ mag. Dugan (1915) conducted a detailed period study of the system and found that the mid-eclipse times show two sinusoidal variations with periods of 10.7 and 26.8 yr, while Rafert (1982) found only one cyclic period of 20.3 yr. However, many mid-eclipse times have been obtained in the past 32 yr. It is necessary to re-analyse the behaviour of the change in the observed period.

In this paper, the $O-C$ data are derived from all available mid-eclipse times in Section 2, where we also present several new data. In Section 3, we develop the fitting procedures described in Yuan \& \c{S}enavc{\i} (2014, hereafter Paper I). In Section 4, we test the Keplerian model, and find that some of the best-fit elements are valid. Finally, we summarize our results and give our conclusions in Section 5.

\section{Eclipse-timing variations}
\label{sect:ec-ti}

\renewcommand{\thefootnote}{\arabic{footnote}}
We carried out the CCD observations of Z Dra in 2014 April and 2015 February using the 40-cm Schmidt-Cassegrain telescope at the Ankara University Kreiken Observatory of Turkey (AUKR-T40), and the 60-cm Cassegrain telescopes at Yunnan Observatory (YNAO-60) in China. The exposure times we adopted in 2014 April are 60 s, 30 s, 20 s, and 15 s in $B$, $V$, $R$, and $I$ bands, respectively. The exposure times in 2015 February is 80 s in the $V$ band and 50 \emph{}s in the $R$ band. The comparison, check stars are BD+72$^{\circ}$545 ($\alpha_{J2000.0}$ = $11^{h}45^{m}56.^{s}1$, $\delta_{J2000.0}$ = $72^{\circ}05^{\prime}44.^{\prime\prime}5$) and GSC 4395-201 ($\alpha_{J2000.0}$ = $11^{h}43^{m}21.^{s}5$, $\delta_{J2000.0}$ = $72^{\circ}06^{\prime}34.^{\prime\prime}2$), respectively. The data reduction is performed by using the aperture photometry
package \textsc{iraf}{\footnote[1]{\textsc{iraf} is developed by the National
Optical Astronomy Observatories, which are operated by the Association of Universities for Research in Astronomy, Inc., under contract to the National Science Foundation.}} (bias subtraction, flat-field division). Extinction corrections are ignored as the comparison star is very close to the variable. We fit the transit center of the eclipse by using the technique of Kwee \& van Woerden (1956). In total, 3
new mid-eclipse times are obtained and listed in Table 1.

\begin{table*}
\begin{minipage}{12cm}
\caption{Several new mid-eclipse times of Z Dra.}
\begin{tabular}{cccccc}\hline
HJD (UTC) & BJD (TDB) & Errors & Min. & Filters &  Origin\\
2400000+     &  2400000+   &  (d)  &  &   & \\\hline
56775.4605 & 56775.46127 & $\pm0.0002$ &  I  &  $B$ &   AUKR-T40\\
56775.4605 & 56775.46127  & $\pm0.0002$ &  I  &  $V$ &  AUKR-T40\\
56775.4602 & 56775.46097  & $\pm0.0002$ &  I  &  $R$ &  AUKR-T40\\
56775.4603 & 56775.46107  & $\pm0.0002$ &  I  &  $I$ &  AUKR-T40\\
57063.2409 & 57063.24167  & $\pm0.0002$ &  I  &  $V$ &  YNAO-60\\
57063.2410 & 57063.24177  & $\pm0.0002$ &  I  &  $R$ &  YNAO-60\\
57071.3858 & 57071.38657  & $\pm0.0002$ &  I  &  $V$ &  YNAO-60\\
57071.3860 & 57071.38677  & $\pm0.0002$ &  I  &  $R$ &  YNAO-60\\
\hline
\end{tabular}
\end{minipage}
\end{table*}

The Lichtenknecker Database of the BAV{\footnote[2]{http://www.bav-astro.de/index.php?sprache=en}} and the O-C Gateway Database{\footnote[3]{http://var.astro.cz/ocgate/}} list all available mid-eclipse times of Z Dra in the literature. In addition, 15 mid-eclipse times between 1928 and 1949 were obtained by Kreiner, Kim \&
Nha (2001) and kindly sent to us (private communication). Three visual and photographic times (HJD 2415787.7856,
2451728.4900, and 2453209.4470) are discarded for their large deviation from the $O-C$ curve. In total, we have collected 820 mid-eclipse times, which have a time span of 125 yr. All the data are plotted in Fig. 1.

Most mid-eclipse times were published without uncertainties. Therefore, a probable uncertainty of $\sigma=\pm0.0003$ d is assumed for the photoelectric and CCD data, and $\pm0.005$ d for the photographic, plate and visual data. Considering the CCD times obtained simultaneously in different filters may differ from each other by as large as $\pm0.0003$ d, the uncertainty of $\pm0.0003$ d is adopted if the mid-eclipse time is obtained in a single filter and with uncertainty less than $\pm0.0003$ d. Eventually, all high-precision (i.e., $\sigma < 0.001$ d) data spread over the last 16 yr, and most low-precision (i.e., $\sigma > 0.001$ d) data over the remaining time.

Since the Heliocentric Julian Dates (HJD) in Coordinated Universal Time (UTC)
system are not uniform, all eclipse times after 1950 have been converted to Barycentric Julian Dates (BJD) in Barycentric Dynamical Time (TDB) system using the UTC2BJD{\footnote[3]{http://astroutils.astronomy.ohio-state.edu/time/}}
procedure provided by Eastman, Siverd \& Gaudi (2010). For the eclipse times before 1950, the relation between the Universal Time (UT) and the Terrestrial Time (TT) given by Duffett-Smith \& Zwart (2011) is adopted visually for the conversion, producing additional uncertainties of a few seconds, which are much smaller than their assumed uncertainty of 0.005 d (i.e., 432 s).

Based on the eclipse times between 2011 and 2014, a new linear ephemeris
\begin{equation}
{\mathrm{Min}}~I = \mathrm{HJD}2456775.4604 + 1^{d}.35745406 \times{E}
\end{equation}
is obtained for future observations. In this paper, the eclipse-timing residuals, $O-C$, are
computed with respect to the linear ephemeris given by Kreiner, Kim \&
Nha (2001),
\begin{equation}
{\mathrm{Min}}~I = \mathrm{BJD}2443499.7305 + 1^{d}.35743190 \times{E},
\end{equation}
where $E$ denotes the cycle number. The $O-C$ data is displayed in Fig. 1(a).

\begin{figure}
\vspace{0.5cm}
\begin{center}
\includegraphics[width=8.5cm]{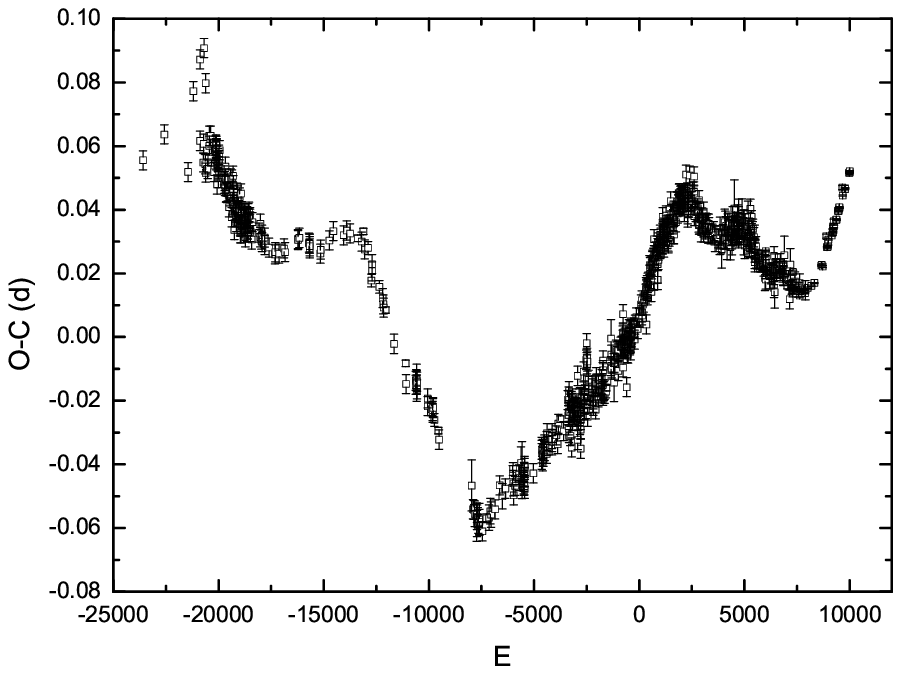}
\caption{The $O-C$ diagram of Z Dra based on the 820 original data. Note that the error bars are smaller than the squares for the high-precision data in the last 4000 cycles.}\label{fig1}
\end{center}
\end{figure}

\section{Data analysis and LTT models}Usually, there is mass transfer between two components in an Algol-type binariy, and the observed period should increase or decrease, suggesting that the $O-C$ curve has a parabolic trend. It is obvious that a single parabola can not describe the $O-C$ curve very well, suggesting that an additional periodic model may be required. Since the data are sampled unevenly with different uncertainties, it is inappropriate to use the parabolic model and the periodic model in turn. If the residuals of a best fit are used for another fit, one would obtain a best-fit solution different from that given by a combination of both models. Therefore, a quadratic plus sinusoidal model
\begin{equation}
O-C = T_{O}(E) - T_{C}(E) = C_0 + C_1\times{E}+ C_2\times{E^2} + A\sin(2{\pi}t/P_{3}) + B\cos(2{\pi}t/P_{3}).
\end{equation}
is used to calculated the generalized Lomb-Scargle (GLS) periodogram, which is plotted in Fig. 2(a).
In equation (3), $A$, $B$, and $C_{0,1,2}$ are free coefficients. In the GLS periodogram, the power peaks at 18.8, 20.5, and 56.0 yr. As pointed out by Zechmeister \& K\"{u}rster (2009), the GLS periodogram can give a good initial guess for the best Keplerian period with only a slight frequency shift.

Then, we use simultaneously a second-order polynomial and one LTT term to fit the $O-C$ values:
\begin{equation}
O-C = T_{O}(E) - T_{C}(E) = C_0 + C_1\times{E}+ C_2\times{E^2} + {\tau}_3,
\end{equation}
where the LTT term ${\tau}_3$ is derived from Keplerian orbits (Irwin 1952), and can be expressed as
\begin{equation}
{\tau}_3=\frac{a_{3}\sin
i_3}{c}\Big[\frac{1-{e_3}^2}{1+e_3\cos\nu_3}\sin(\nu_3+\omega_3)+e_3\sin\omega_3\Big].
\end{equation}
In Equation (5), $a_{3}{\sin}i_{3}$ is the projected semimajor axis of the eclipsing pair around the barycentre of the triple system ($i_{3}$ is the orbital inclination of the companion with respect to the tangent plane of the sky). $e_3$ is the eccentricity, and $\omega_3$
is the argument of the periastron measured from the ascending node in the tangent
plane of the sky. $\nu_3$ is the true anomaly, which is related with the
mean anomaly $M=2\pi(t-T_3)/P_3$, where $T_3$ and $P_3$ are the time of the periastron
passage and orbital period, respectively.

For fixed $e_3$, $T_3$, and$P_3$, $\nu_3$ can be computed for all mid-eclipse times. Then, we fit the $O-C$ data with equation (4), and get the goodness-of-fit statistic, $\chi^2$, which is the weighted sum of the squared difference between the $O-C$ values $y_i$ and the model values $y(t_i)$
at eclipse times $t_i$:
\begin{equation}
\chi^2 = \sum_{i=1}^{N}\big[\frac{y_i-y(t_i)}{{\sigma}_j}\big]^2 = W\sum_{i=1}^{N}w_i[y_i-y(t_i)]^2,
\end{equation}
where
\begin{equation}
w_i =
\frac{1}{W}\frac{1}{{{\sigma}_i}^2},
\end{equation}
and
\begin{equation}
W = \sum_{j=1}^{N}\frac{1}{{{\sigma}_j}^2}.
\end{equation}
In equation (6), $\sigma_i$ is the uncertainties of the $O-C$ data $y_i$, and $N$ the number of data.

Stepping through $e_3$ and $T_3$, we obtain the local $\chi^2$ minimum for the fixed $P_3$, i.e., $\chi^{2}(P_3)$. Since $\sum_{i=1}^{N}w_{i}=1$, $\sqrt{\chi^{2}(P_3)/W}$ can be regarded as the weighted root-mean-square (rms) scatter around the best fit for the fixed $P_3$ (Marsh et al. 2014). After searching $P_3$, the global chi-square minimum, ${\chi}^2_{global}$, can be found. Some local $\chi^2$ minima at $P_{3} > 100$ yr give the companion with a mass more than $200M_{\bigodot}$, and are ruled out by us. Normalized by ${\chi}^2_{global}$, we obtain a power spectrum (Zechmeister \& K\"{u}rster 2009; Cumming et al. 1999; Cumming 2004),
\begin{equation}
p(P_3)\equiv\frac{{\chi}_{0}^{2} - \chi^{2}(P_3)}{{\chi}_{global}^2},
\end{equation}
where the constant ${\chi}_{0}^{2}$ is the best-fit statistic of a fit of a parabola to the data.
Fig. 2(b) shows the one-dimensional Keplerian periodogram as well as the best-fit eccentricity. If the minimum rms scatter, $\sqrt{\chi^{2}_{global}/W}$, is taken as a noise in the power spectrum, and $\sqrt{({\chi}_{0}^{2} - \chi^{2})/W}$ as a signal, the $\sqrt{p(P_3)}$ would be the signal-to-noise ratio (S/N).

\begin{figure}
\vspace{0.5cm}
\begin{center}
{\includegraphics[width=8.6cm]{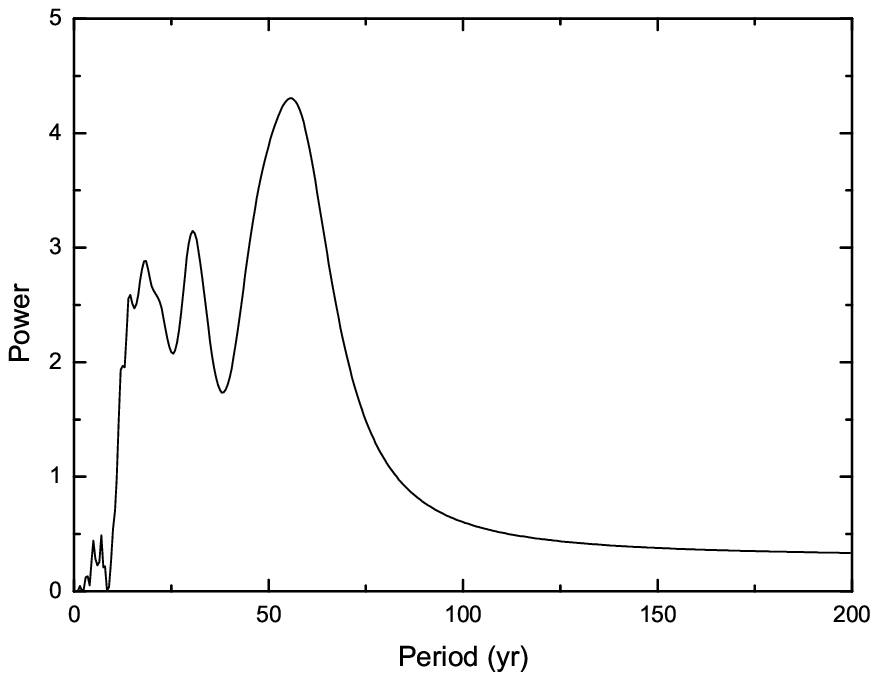}
 \centerline{(a)}
 \includegraphics[width=9.0cm]{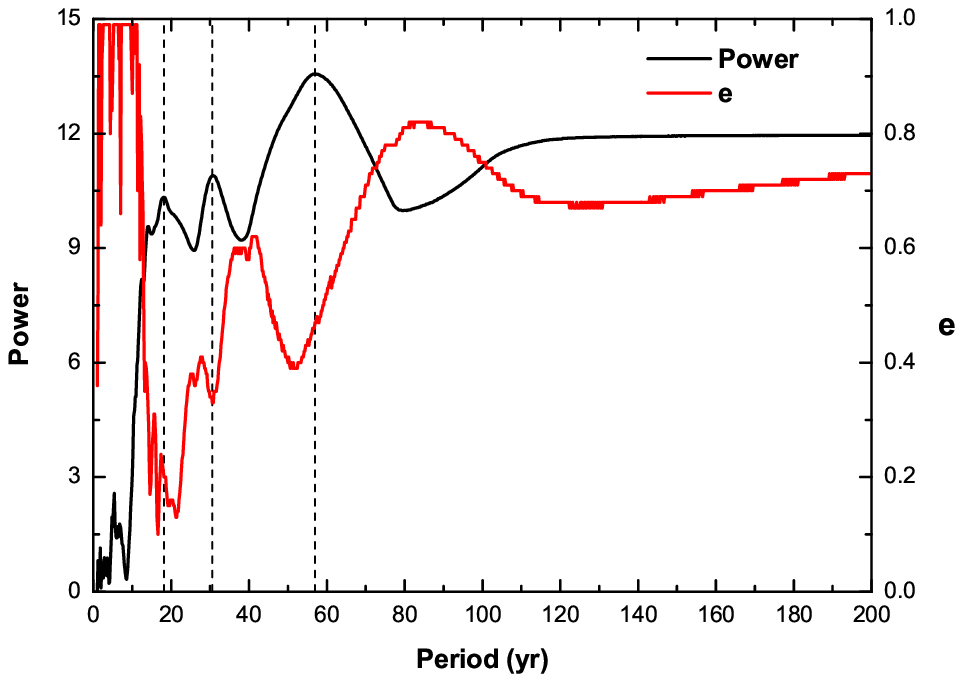}
 \centerline{(b)} }
 \caption{The GLS periodogram (a) and Keplerian periodogram (b) of Z Dra. The dashed vertical lines marks three peaks in the Keplerian periodogram. The red line represents the best-fit eccentricity corresponding to $\chi^{2}(P_3)$.}\label{fig2}
\end{center}
\end{figure}

All one-dimensional periodograms show an extremely significant periodicity at $\sim$60 yr, suggesting a companion with a period of $\sim$60 yr (hereafter, referred to as Z Dra (AB)C). In all periodograms, the power also peaks at $P$ = $\sim$30 yr, suggesting another companion with an orbital period of $\sim$30 yr (hereafter, referred to as Z Dra (AB) D). The companion is in/around a 2:1 mean-motion resonance (MMR) with Z Dra (AB)C. The $\sim$20 yr periodicity reported by Rafert (1982) is obvious in Fig. 2(a). If the $\sim$20 yr signal exists, the eclipsing binary has a third companion (hereafter, Z Dra (AB)E). It is interesting that Z Dra (AB)C, D, and E are in 6:3:2 MMRs. Furthermore, the three maxima in the power spectrum (i.e., three local $\chi^2$ minima) appear near three $e$ minima.

We also note that, due to the short time coverage, the power increases continuously from $\sim90$ yr , but remains all the time below the $\sim60$ yr peak. Although the power at long periods ($P_{3}>90$ yr) is still large, the best-fit solutions at long periods gives an eccentricity larger than 0.70. Such a large eccentricity is physically unlikely. An large eccentricity often implies a large gravitational perturbation from other companions. The statistic, $p(P_3)(N-8)/4$, follows Fisher's $F$ distribution with 4 and $N-8$ degrees of freedom (Bevington \& Robinson 1992). Integrating the distribution function and multiplying it by the number of independent frequencies gives the false alarm probability (FAP) less than $10^{-30}$ for the three peaks (Cumming et al. 1999; Cumming 2004). In fact, the FAP values should be derived from a suitable model. But, the one-companion model is not suitable for the Z Dra system (see below).

The best fits corresponding to the 60 yr periodicity are plotted in Fig 3(a), and listed in the second column (Solution 1) of Table 2. For safety, we also use a third-order polynomial instead of the second-order polynomial in Equation (4), and obtain Solution 2, which is shown in Fig 3(b). As shown in the bottom panels of Figs 3(a) and 3(b), the residuals at $\sim$BJD2446000 reach as large as 0.02 d, which are much larger than their uncertainties. It seems that the residuals show cyclic variation with a period of $\sim$30 yr.

\begin{figure}
\vspace{0.5cm}
\begin{center}
{\includegraphics[width=8.5cm]{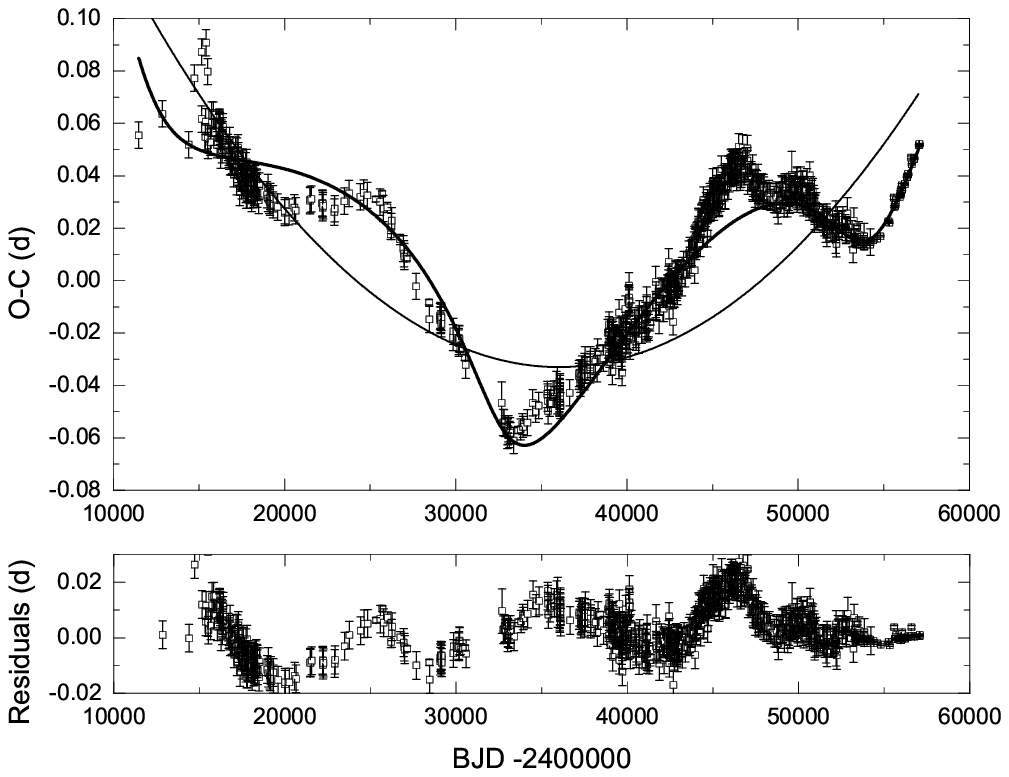}
 \centerline{(a)}
 \includegraphics[width=8.5cm]{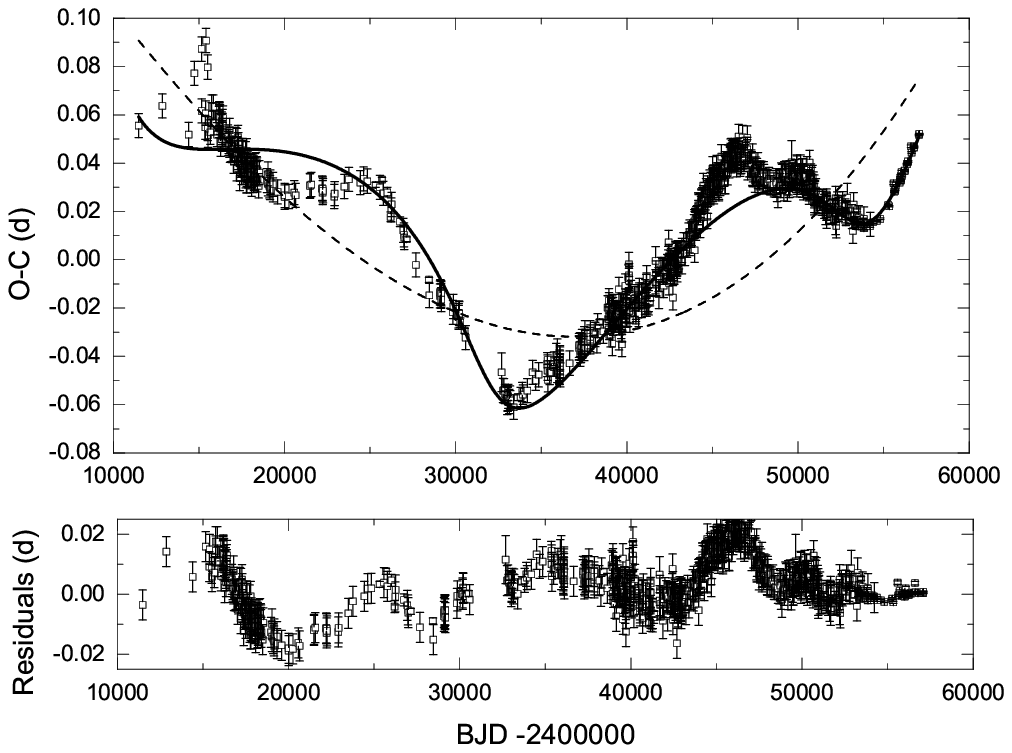}
 \centerline{(b)}}
 \caption{The one-companion fit to the eclipse-timing variations of Z Dra. (a) The overplotted solid line denotes the best fit with Equation (3), and the dashed line only represents the second-order polynomial in the ephemeris. The residuals of the best fit are displayed in the lower panel. (b) The same as figure (a) but a three-order polynomial is adopted. }\label{fig3}
\end{center}
\end{figure}

To determine further whether there are two periodicities in the $O-C$ data, we also use a second-order polynomial plus two-LTT ephemeris to fit the $O-C$ data. We search for the best period in 40-90 yr with one LTT term, and the other LTT term in 10-40 yr. The linearized Keplerian fitting method (Beuermann et al. 2012; Paper I), which is very similar to the method of the one-dimensional periodogram above, is used to calculate a two-dimensional periodogram. The least-squares fit to the 820 data involves thirteen free parameters, three for the second-order polynomial in the ephemeris, and five orbital elements ($P_k$, $e_k$, ${\omega}_k$, $T_k$, and $a_{k}{\sin}i_{k}/c$) for each companion. If all parameters are free, the number of degrees of freedom (DOF) is therefore 807. The constraints on the two orbital periods are shown in Fig. 4(a). The ${\chi}^{2}$ contour levels of 1.05, 1.2, 1.5, 2.0, 3.0, 4.0 and 5.0 have been normalized by division of the global chi-square minimum, ${\chi}^2_{global}$. In addition, the best-fit eccentricities of Z Dra (AB)C and D are plotted in Fig. 4 (b) and (c), respectively. In the two-dimensional periodogram, the global $\chi^2$ minima at ($P_4\simeq60$ yr, $P_3\simeq30$ yr) confirm Z Dra (AB)C and D, and the local $\chi^2$ minima at ($P_4\simeq60$ yr, $P_3\simeq20$ yr) reveals Z Dra (AB)C and E. Both $\chi^2$ minima lie close to the points of the $e_3$ minima and also the $e_4$ minima.

\begin{figure}
\vspace{0.5cm}
\begin{center}
{\includegraphics[width=8.0cm]{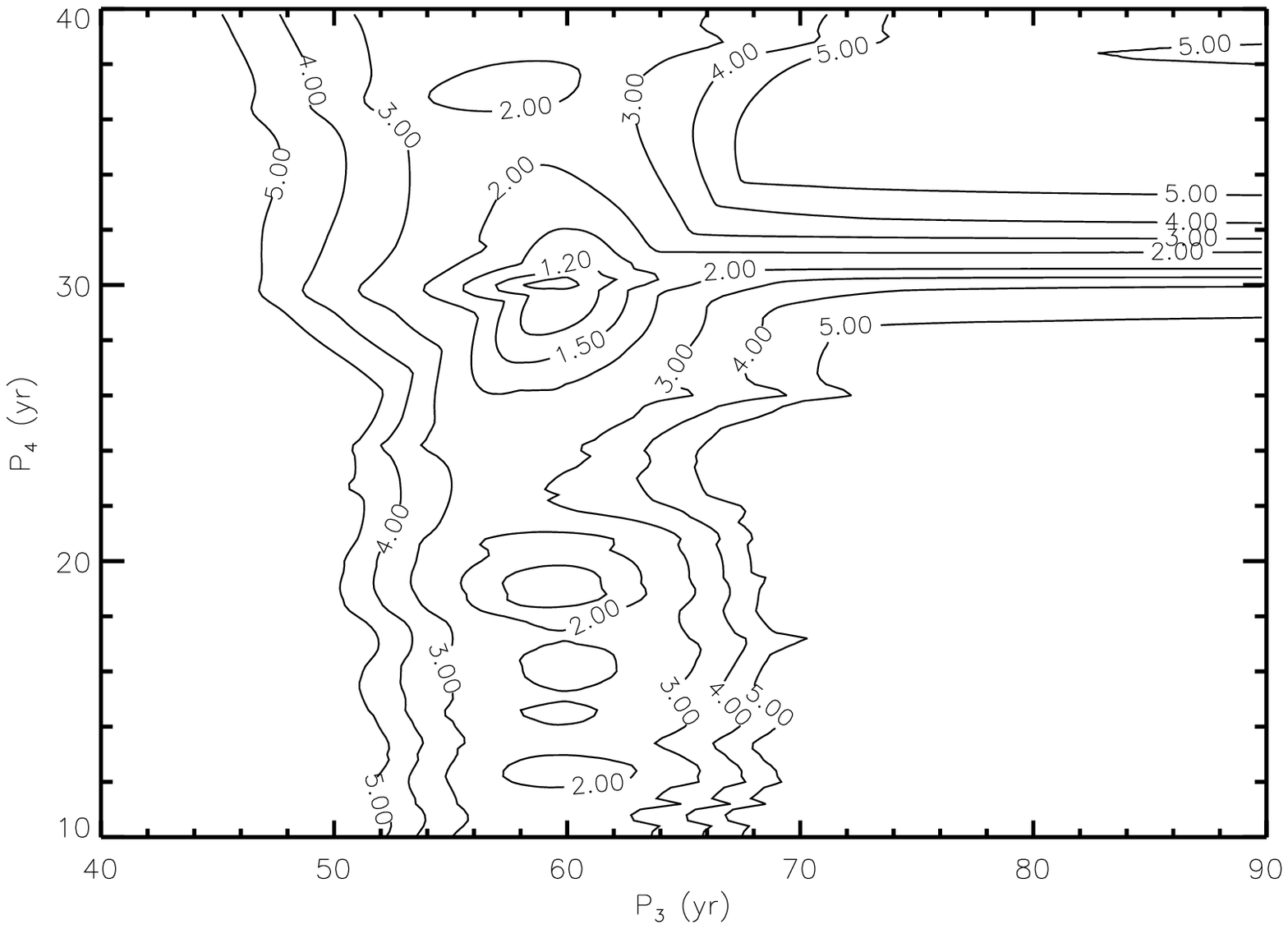}
 \centerline{(a)}
 \includegraphics[width=8.0cm]{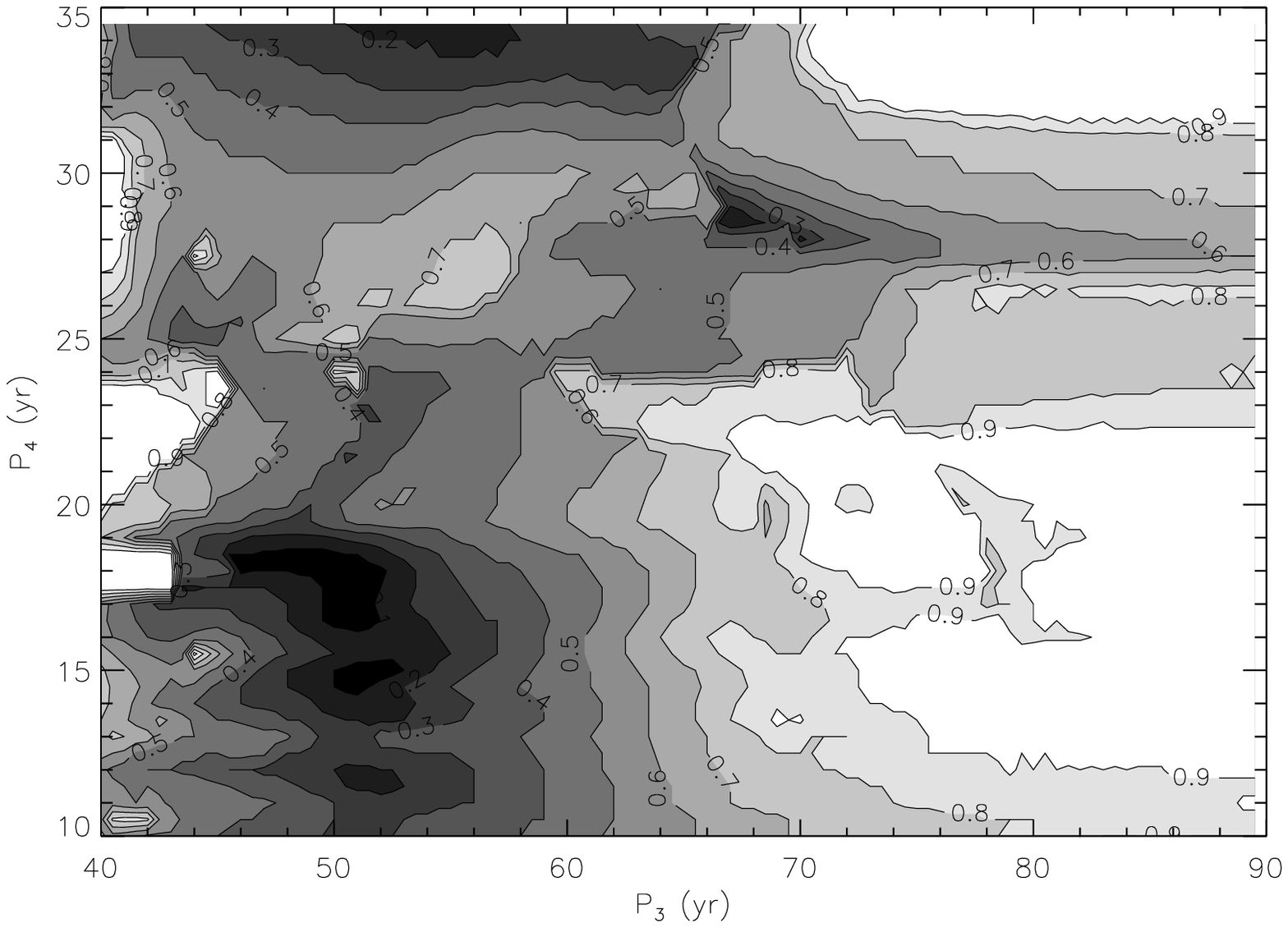}
 \centerline{(b)}
  \includegraphics[width=8.0cm]{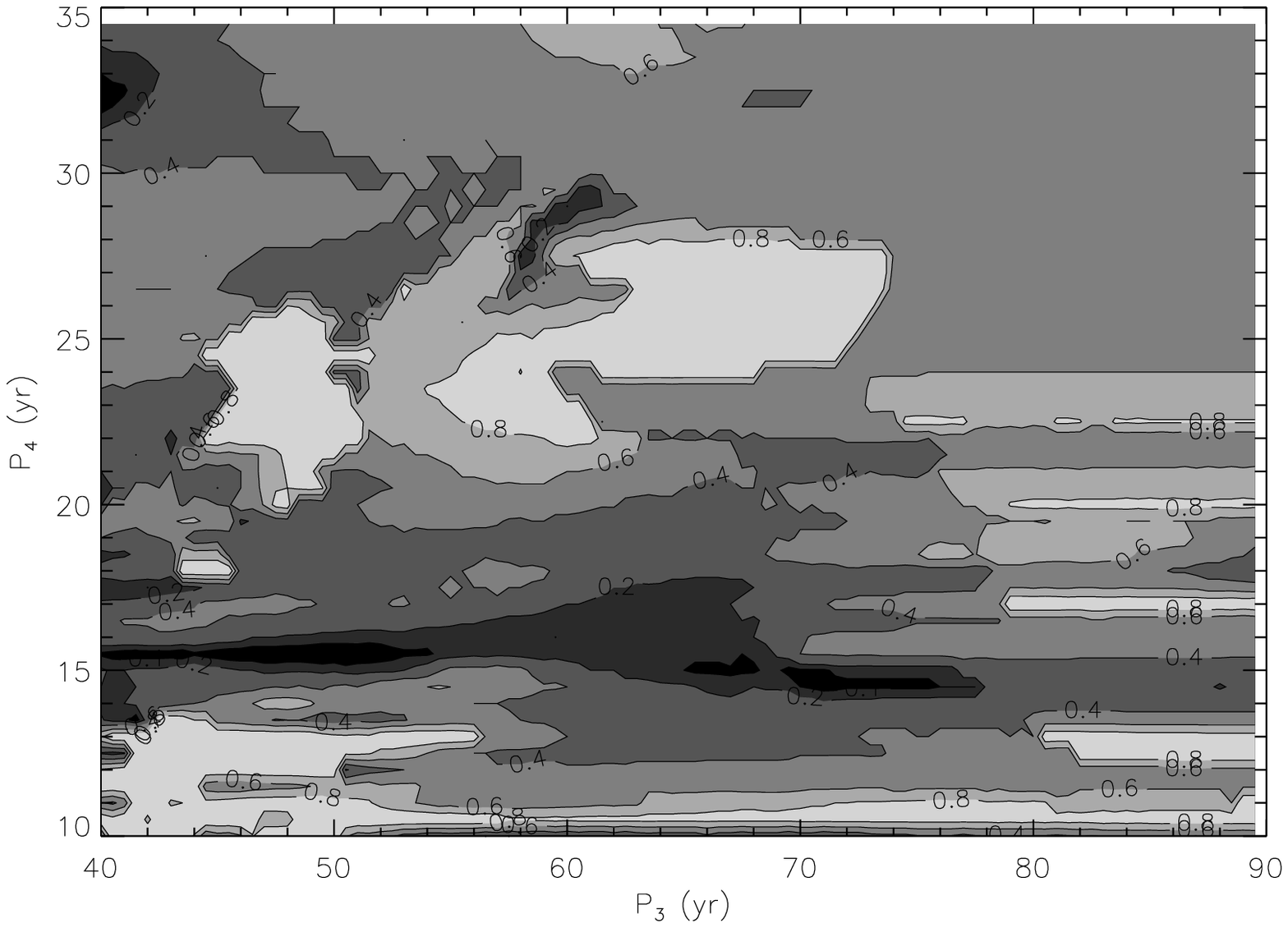}
 \centerline{(c)} }
 \caption{(a) Two-dimensional Keplerian periodogram derived from a second-order polynomial plus two-LTT model. The $\chi^2$ contours of 1.05, 1.2, 1.5, 2.0, 3.0, 4.0 and 5.0 have been normalized by division of the global ${\chi}^{2}$ minimum. (b): The best-fit eccentricity ($e_3$) of Z Dra (AB)C as a function of ($P_3$,$P_4$). The darker the region is, the smaller the eccentricity is. (c): the same as figure (b) but for $e_4$.}\label{fig4}
\end{center}
\end{figure}

Based on the best solution in the two-dimensional periodogram, the Levenberg-Marquardt fitting algorithm (Markwardt 2009) is adopted to search for the improved solutions. The improved fits are plotted in Figs 5(a) and 5(b). The corresponding parameters and $\chi^2$ are listed in the fourth and fifth columns (i.e., Solutions 3 and 4) of Table 2. After the parabolic trend is removed, the residuals are displayed in Fig. 6, where two sets of periodic variations can be seen more clearly. Compared to Solution (1), the reduced ${\chi}^{2}_{\nu}=2.6$ (${\chi}^{2} = 2118.3$ for 807 DOF) in Solution (3) improves greatly, but is yet unacceptable (Bradt 2004). The large $\chi^2_{\nu}$ is due to the large uncertainties of the old $O-C$ data before BJD2415700 (i.e., $E < -20500$), and perhaps a third set of cyclic variation in the residuals.

\begin{figure}
\vspace{0.5cm}
\begin{center}
{\includegraphics[width=8.0cm]{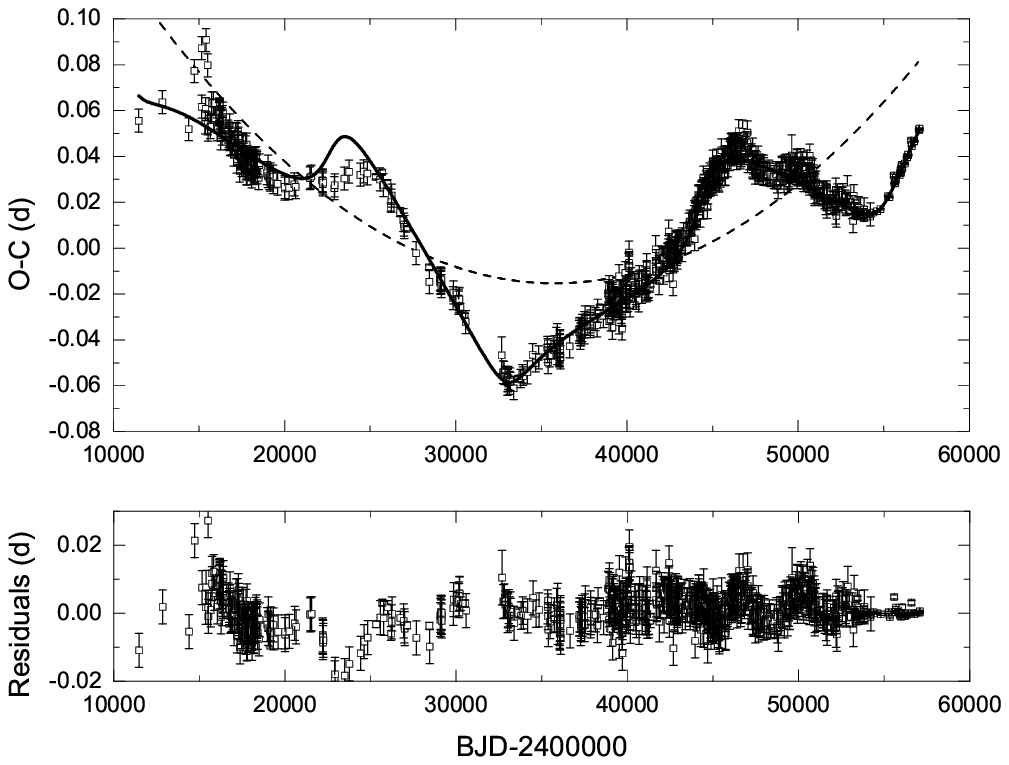}
 \centerline{(a)}
 \includegraphics[width=8.0cm]{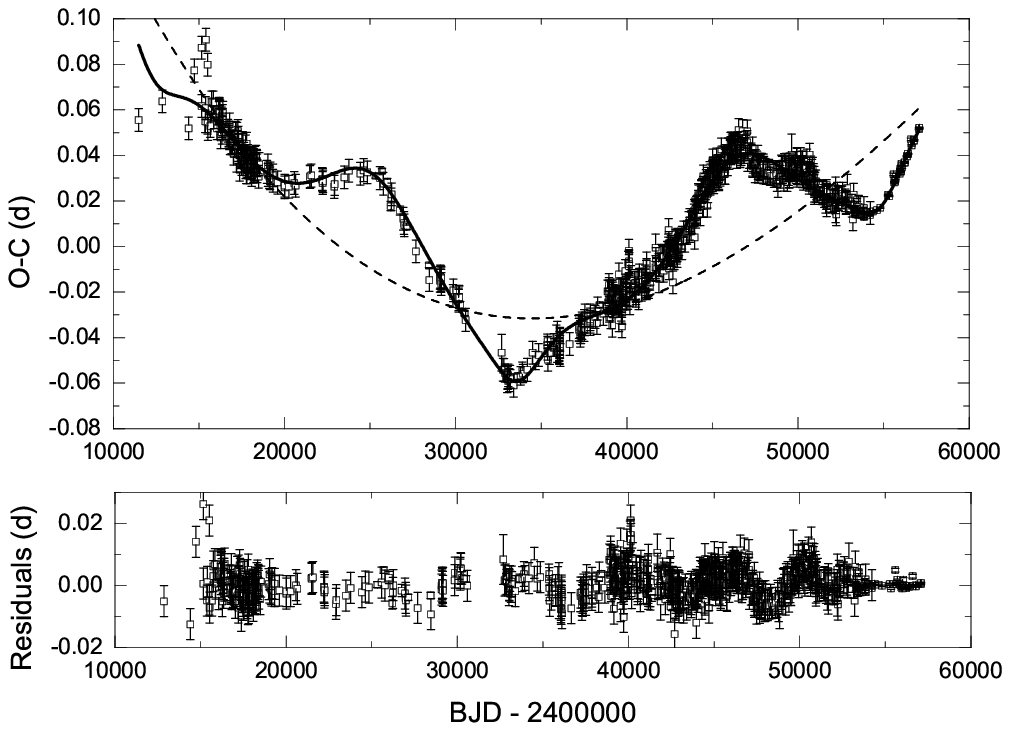}
 \centerline{(b)}}
 \caption{The two-companion fit to the eclipse-timing variations of Z Dra when a second-order polynomial trend (a) or a third-order polynomial trend (b) is considered. The residuals of the best fit are displayed in the lower panel of each figure. The overplotted solid line denotes the best fit with a polynomial plus two-LTT ephemeris, and the dashed line only represents the polynomial in the ephemeris.}\label{fig5}
\end{center}
\end{figure}

\begin{figure}
\vspace{0.5cm}
\begin{center}
\includegraphics[width=8.5cm]{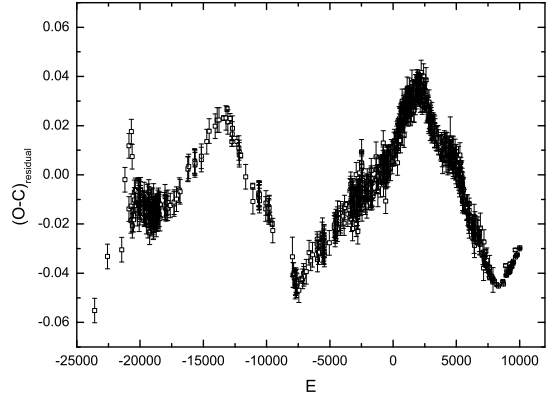}
\caption{The same as Fig. 1, but subtracted by the parabolic trend given by Solution (3).}\label{fig1}
\end{center}
\end{figure}

\begin{table*}
\caption{The best-fit parameters for the LTT orbits of Z Dra.}
\begin{tabular}{lcccc}\hline
parameter & Solution 1 & Solution 2 & Solution 3   &  Solution 4  \\\hline
$C_0$ (d) & -0.0197$\pm0.0001$ & -0.0211$\pm0.0010$ & -0.0025$\pm0.0007$ & -0.0146$\pm0.0009$ \\
$C_1$ ($\times10^{-6}$ d) & 4.80$\pm0.01$ & 4.44$\pm0.03$ & 4.49$\pm0.02$ & 4.88$\pm0.07$ \\
$C_2$ ($\times10^{-10}$ d) & 4.35$\pm0.01$ & 4.80$\pm0.20$ & 3.94$\pm0.01$ & $ 3.16\pm0.03$ \\
$C_3$ ($\times10^{-15}$ d) & & -3.85$\pm0.25$& & -5.06$\pm0.30$\\\hline
$P_4$ (yr)  &   &   & 29.81$\pm0.08$ & 29.05$\pm0.08$ \\
$T_4$ (BJD) &   &   & 2400464$\pm164$ & 2403688$\pm513$ \\
$e_4$       &   &   &  0.43$\pm0.01$ &  0.11$\pm0.03$   \\
$a_4\sin i_4$ (au)& & & 2.14$\pm0.03$ & 1.92$\pm0.09$ \\
$\omega_4 (deg)$  & & & 285.6$\pm3.9$ & 83.1$\pm26.0$ \\
$m_{4}$ ($M_{\bigodot}$, $i_4 = 90^{\circ}$) & & & 0.39$\pm0.03$ & 0.33$\pm0.04$ \\
$A_{4}$ (au, $i_4 = 90^{\circ}$) & & & 12.74$\pm0.3$ & 12.3$\pm0.2$ \\\hline
$P_3$ (yr) & 57.49$\pm0.27$ & 59.11$\pm0.15$ & 59.41$\pm0.12$ & 58.07$\pm0.12$ \\
$T_3$ (BJD) & 2411634$\pm77$ & 2410320$\pm127$ & 2401400$\pm109$ & 2412114$\pm101$ \\
$e_3$ &  0.42$\pm0.01$ & 0.41$\pm0.01$ & 0.62$\pm0.02$ & 0.56$\pm0.01$ \\
$a_3\sin i_3$ (au) & 5.52$\pm0.06$ & 5.79$\pm0.07$ & 6.05$\pm0.07$ & 5.61$\pm0.06$ \\
$\omega_3 (deg)$ & 232.9$\pm1.2$ & 226.4$\pm1.1$ & 76.8$\pm4.0$ & 240.5$\pm1.6$ \\
$m_{3}$ ($M_{\bigodot}$, $i_3 = 90^{\circ}$) & 0.70$\pm0.01$ & 0.73$\pm0.01$ & 0.77$\pm0.02$ & 0.77$\pm0.03$ \\
$A_{3}$ (au, $i_3 = 90^{\circ}$) & 20.5$\pm0.3$ & 20.9$\pm0.4$ & 22.3$\pm0.3$ & 21.9$\pm0.2$ \\\hline
$\chi^2$ & 6221.1 & 6110.6 & 2118.3 & 2005.8 \\\hline
\end{tabular}
\end{table*}

As shown in Figs 5(a) and 5(b), the LTT signal of Z Dra (AB)E can be seen in the residuals of the two-companion fit. Further fit reveals that Z Dra (AB)E has an orbital period of $P_5=\sim 20.1$ yr and a mass of $\sim0.2M_{\bigodot}$. Z Dra (AB)E produces a cyclic $O-C$ variation with a semi-amplitude of $a_5\sin i_5=\sim 0.8$ au, which is much smaller than $a_3\sin i_{3}$. In such a case, it is also possible that such small signal arises from unavoidable and slight imperfection in the double-Keplerian model (see below).

\section{Tests of the so-called Keplerian model}
\label{sect:test}

Based on an assumed inclination for one companion, its mass ($m_k$) can be estimated from the following mass functions
\begin{align}
& \frac{(m_4{\sin}i_4)^3}{(m_{b} + m_4)^2} = \frac{4\pi^2}{G{P_4}^2}\times(a_4\sin
i_4)^3,\\
& \frac{(m_3{\sin}i_3)^3}{(m_{b} + m_4 + m_3)^2} = \frac{4\pi^2}{G{P_3}^2}\times(a_3\sin
i_3)^3,
\end{align}
where $G$ is the Newtonian gravitational constant. For simplicity, the central eclipsing binary is treated as a single object ($m_b$) with a mass equal to the sum of the masses of both components. In the case of Z Dra, $m_b = 1.90 M_{\bigodot}$ (Terrell 2006). It is important to keep in mind that the $m_4$ and $m_3$ derived in this way are just approximate masses since the mass functions are derived from Kepler's third law. If the orbital inclinations of both companions are $90.0^{\circ}$, the outer companion Z Dra (AB)C has the minimum mass of $\sim0.8~M_{\bigodot}$, whereas the internal companion Z Dra (AB)D are M dwarfs with masses of $\sim0.4~M_{\bigodot}$. It is obvious that Z Dra is a general N-body system. Given $m_4$ and $m_3$, we can calculate the semimajor axes of the two companions by the equations, $A_{3}=a_3 \cdot (m_b+m_3)/m_3$ and $A_{4}=a_4 \cdot (m_b+m_3+m_4)/m_4$. The minimum $A_{4}$ are about $420$ times larger than the separation between Z Dra A and B ( $6.38R_{\bigodot}=0.030$ AU), suggesting that the central eclipsing pair can be treated as a single object.

Assuming that Z Dra (AB) C and D revolve around Z Dra AB in coplanar Keplerian orbits with $i_{3}=i_{4}=90.0^{\circ}$, the centripetal force ($F_{c}$) of Z Dra (AB)D from the eclipsing pair is compared with the gravitational perturbation ($F_{p}$) from the outer companion, Z Dra (AB)C. The "relative perturbation", $F_{p}/F_{c}$, is calculated on a 130 yr timescale (BJD 2410000 - 2457482). In the process of the calculation, we calculate the coordinates of the triple objects. Then, the forces of gravity are derived from their separations and masses.

For Solution (3) or (4), the result reveals that $F_{p}/F_{c}$ peaks at $\sim0.25$ with a mean value of $\sim0.09$ (see Fig. (7)). The gravitational perturbation will decrease if the errors of the orbital parameters, especially $\omega_{3,4}$ and $T_{3,4}$, are considered. Although mutually tilted orbits can reduce the gravitational perturbation, the mutual perturbation between the two companions can not be neglected. The Keplerian formulae serves only as a convenient, mathematical description of the $O-C$ data.

\begin{figure}
\vspace{0.5cm}
\begin{center}
\includegraphics[width=8cm]{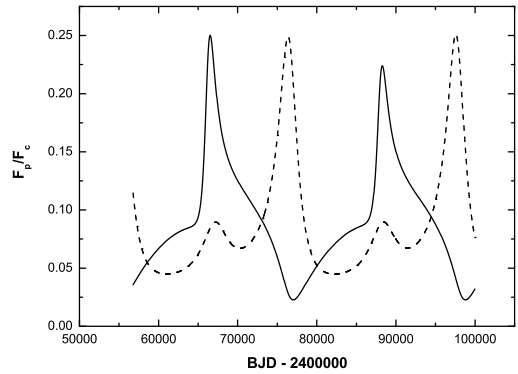}
\caption{The ratio of the gravitational perturbations from the outer companion (Z Dra (AB)C) to centripetal forces from the central eclipsing pair (Z Dra A and B), both of which act on Z Dra (AB)D in opposition to each other. The solid line is derived from Solution (3), and the dashed line from Solution (4).} \label{fig6}
\end{center}
\end{figure}

Generally, the Newtonian LTT signals derived from N-body simulations differ more or less from those given by the double-Keplerian model (Marsh et al. 2014; Go\'{a}dziewski et al. 2012; Go\'{a}dziewski et al. 2015). If we are only interested in the orbital periods of two comapanions, the LTT value caused by the outer companion, $(O-C)_3$, can be still fitted by the LTT model given by Equation (4). In this case, the best-fit parameters have no physical meaning except for the orbital period and the projected semimajor axis. (Strictly, $a_{3}{\sin}i_{3}$ is an half of the width of the orbit in the line-of-sight direction). Comparing with a sinusoidal model with three free parameters (i.e., $A{\sin}(Bt + C)$), the LTT model has five free parameters, and can present more complex and abundant $O-C$ curves (see Fig. 8). There must be a Keplerian $O-C$ curve whose shape is the most similar to the true $(O-C)_3$. The best-fit Keplerian curve may differ slightly from the true $(O-C)_3$. Note that the true $O-C$ value is equal to $(O-C)_{3}$ plus $(O-C)_{4}$ if the parabolic trend is neglected. Such slight deviations would have some influence on a second fit to $(O-C)_4$, which is caused by the inner companion, In Solutions (3) and (4), $a_4\sin i_{4}$ is about one third of $a_3\sin i_{3}$, suggesting that the influence on the second fit is also at a low level. Therefore, the result of the 2:1 MMR is valid, and the masses of Z Dra (AB)C and D are approximate. As for the low-amplitude ($a_5\sin i_5=\sim 0.8$ au) variation, it may arise from slight imperfection in the double-Keplerian model.

\begin{figure}
\vspace{0.5cm}
\begin{center}
\includegraphics[width=9cm]{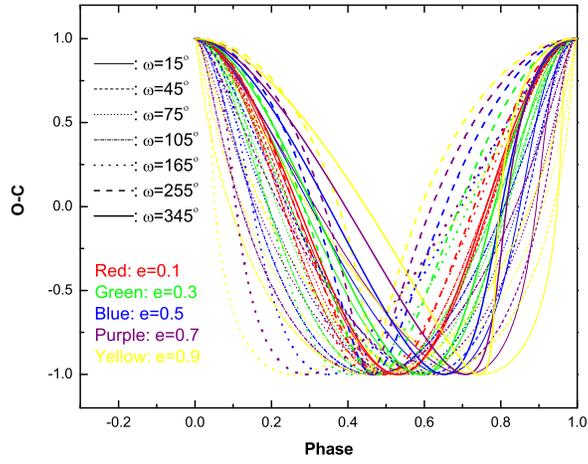}
\caption{All kinds of $O-C$ curves presented by the so-called Keplerian model given by Equation (4). Different colors refer to different eccentricities, and different line styles and thickness denote the $O-C$ curves with different $\omega$ values. The semi-amplitudes of all $O-C$ curves are normalized to unity. The orbital phase is proportional to the time, and has been shifted so that phase zero corresponds to the $BJD$ time of the $O-C$ maximum.} \label{fig7}
\end{center}
\end{figure}

Finally, we like to remind the reader that the best-fit eccentricities is not equal to the orbital eccentricities accurately. Actually, the observed eccentricity results from the true orbital eccentricity and the deviation of the angular velocity from that of a Keplerian motion. The deviation of the angular velocity arises from the gravitational perturbation from other companions, and therefore should be small since the gravitational perturbation should be small in a stable system. On the other hand, the orbital eccentricity should be also small since an companion with small orbital eccentricity often experience weak gravitational perturbation from other companions. These may be the reason why the $\chi^2$ minima lie close to points of $e_{3,4}$ minima.

\section{Discussions and Conclusions}
\label{sect:disc}

Detailed $O-C$ analyses of Z Dra are performed by using all available mid-eclipse times in the literature as well as eight new mid-eclipse times obtained in this paper. The $O-C$ diagram shows a quadratic or cubic trend. A companion with orbital period more than twice as long as the time window of observation can produce a quadratic/cubic $O-C$ curve, which is actually a section of a cyclic $O-C$ curve. But, the quadratic/cubic trend is often explained by mass transfer between two components. The quadratic trend in Solution (3) represents an observed period increase with a rate of $dP/dt = 2.1 \times 10^{-7} \mathrm{d~yr^{-1}}$, which is a typical value for many contact binary stars (see e.g., Qian 2001, 2003, 2008). The cubic trend in Solution (4) suggests that the observed period increase at a decreasing rate. The mass transfer from the secondary component to the primary will cease in 197 years. Then, the mass will be transferred from the primary component to the secondary one. Comparing with binary evolutionary timescales, a timescale of a few hundreds of years is infinitely short. The mass transfer rate in the eclipsing pair should change little or remain constant over a few hundreds of years, suggesting a quadratic trend rather than a cubic trend. Furthermore, the cubic model does not have a significant advantage over the quadratic model. The best-fit cubic trend in Fig 5(b) is close to the quadratic trend in Fig 5(a). The similar periodicities and $\chi^2$ are obtained in Solutions (3) and (4).

We have searched the $O-C$ data for periodicities. The $O-C$ data show two or more sets of cyclic variations with periods of $59.4$, $29.8$, and possible $\sim20.1$ and $> 80$ yr, suggesting two or more companions around the eclipsing binary. Although we donot know the exact number of companions, there must be more than one companions. If only one companion revolves around the eclipsing binary, and therefore moves in a Keplerian orbit, the single-Keplerian models would fit the $O-C$ data very well. But, Figs 3(a) and (b) show that the single-Keplerian models fail. The two-dimensional periodogram reveals that the companions Z Dra (AB)C and D with periods of the $59.4$ and $29.8$ yr are a most likely combination. As for the long period ($> 80$ yr), Figs 4(b) and (c) show any long-period companion has large $e_3$ and $e_4$, which lie far from the points of the $e_3$ and $e_4$ minima, respectively. Such large eccentricities are physically unlikely.

Although magnetic activity can explain the cyclic variations in the O-C diagram (Applegate 1992; Yuan \& Qian 2007), they are unlikely to produce two/three sets of variations with commensurate periods. The more plausible reason for such variations is the reflex motion of the eclipsing pair induced by two/three companions in a 2:1 or 6:3:2 MMR. In Paper I, two companions were found to be in a near 3:1 MMR orbits around the eclipsing binary SW Lac with periods of 27.0 and 82.6 yr. Both Z Dra and SW Lac have the most numerous mid-eclipsing times, which show complex variations. Perhaps, MMRs are common in such three-body systems.

More than 160 planetary systems have been confirmed so far. About 30\% of them are close to MMRs, particularly near the first order MMRs of 2:1 and 3:2 (Zhang et al. 2014). Furthermore, Beuermann, Dreizler \& Hessman (2013) found that two planetary companions are in a near 2:1 MMR orbits around the eclipsing binary NN Ser. But different from these planetary systems, Z Dra and SW Lac are general three-body systems if the central eclipsing binary is treated as a single object. Our discoveries will help us understand the orbital properties of such three-body systems. Perhaps, MMRs are common in such N-body systems.

In this paper, we have checked the mutual perturbations between Z Dra (AB)C and D, but are unable to test the dynamical stability for three reasons: (1) the so-called double-Keplerian model gives only convenient approximation to the $O-C$ curve, but can not provide exact orbits and correct initial conditions (such as, coordinates, velocities, and masses) for the N-body system. (2) the errors of the orbital parameters should be considered in our dynamic simulations; (3) the inclinations ($i_3$ and $i_4$) and the angle between their ascending nodes in the sky plane ($\theta$) are unknown. The orbitally angular configuration of the outer companion relative to the inner one is determined by $i_3$, $i_4$, and $\theta$ (see Fig. 9). It seems extremely difficult to test the dynamical stability since stable configurations are likely confined to tiny regions of the parameter space for the general three-body system, and all initial conditions must be very accurate in the dynamical analyses.

\begin{figure}
\vspace{0.5cm}
\begin{center}
\includegraphics[width=8cm]{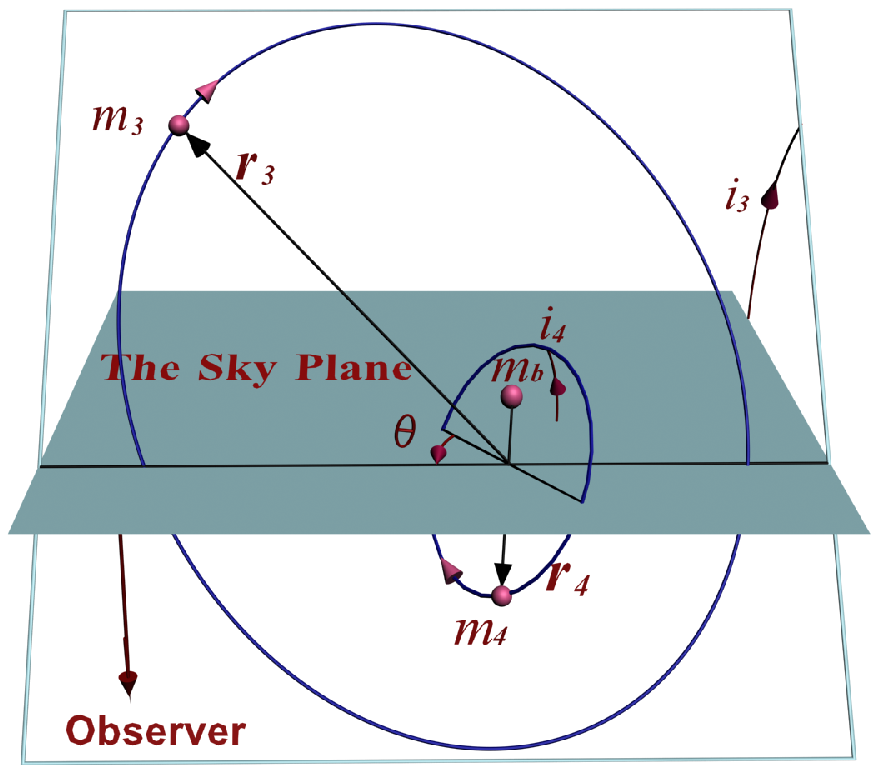}
\caption{Schematic positions of the eclipsing binary ($m_b$) and its two companions ($m_{3,4}$) with respect to the tangent plane of the sky (Paper I).  $i_{3,4}$ denote the orbital inclinations of $m_{3,4}$, respectively. $r_{3,4}$ refer to Jacobian coordinates of $m_{3,4}$. Note that the subscripts `$3$' and `$4$' are assigned to the outer and inner companions, respectively.} \label{fig8}
\end{center}
\end{figure}

We note that the N-body model was used to fit the LTT data of HU Aqr (Go\'{a}dziewski et al. 2012; Go\'{a}dziewski et al. 2015) and NN Ser (Marsh et al. 2014), both of which host two circumbinary planets. In this model, synthetic LTT signals at all epoches are determined through numerical N-body integration, and then compared to the true LTT signals. Based on the reduced ${\chi}^{2}_{\nu}$ and dynamical stability, one can find the best solution. In the case of Z Dra, the masses of Z Dra (AB)C and D are relatively large. We have to fit the LTT data with the masses, velocities and coordinates as free parameters, by integrating the equations of motion (K. Go\'{a}dziewski, private communication). For a general three-dimension system, seven free parameters is need for each companion, giving 17 free parameters in the model. Besides a computational challenge, a possible drawback of this model is that an undiscovered companion can make the N-body integration meaningless. If the orbital period of the companion is more than twice as long as the time window of observation, the LTT signals caused by the companion would show a parabolic/cubic trend in the time window rather than periodic variation. In this case, one would miss the third companion, and therefore its dynamical perturbation. The long-period companion, however, has little influence on any analytic model including a second-order or third-order polynomial, such as the three-order polynomial plus double-Keplerian ephemeris.

As shown in Figs (2) and (4), all periodograms can not provide tight constraints on the periodicity of $>70$ yr. This is mainly attributed to the low precision of the old data and the short time coverage of the $O-C$ data. We therefore encourage follow-up observations of this system to obtain more mid-eclipse times covering as long baseline as possible.

\acknowledgments{We would like to thank Kreiner J. M. for his data and the anonymous referees for some constructive suggestions. This research has also made use of the Lichtenknecker-Database of the BAV, operated by the Bundesdeutsche Arbeitsgemeinschaft f\"{u}r Ver\"{a}nderliche Sterne e.V. (BAV). The computations were carried out at National Supercomputer Center in Tianjin, and the calculations were performed on TianHe-1(A). This work is supported by the National Natural Science Foundation of China (NSFC) (No. U1231121) and the research fund of Ankara University (BAP) through the project 13B4240006.}


\begin{references}
%
\reference{ } Applegate, J. H. 1992, ApJ, 385, 621
%
\reference{ } Bevington, P. R., \& Robinson, D. K. 1992, Data Reduction and Error Analysis for the Physical Sciences (2d ed.; New York: McGraw-Hill)
%
\reference{ } Beuermann, K., Hessman, F. V., Dreizler, S., et al. 2010, A\&A, 521, L60
%
\reference{ } Beuermann, K., Dreizler, S., Hessman, F. V., \& Deller, J. 2012, A\&A, 543, 138
%
\reference{ } Beuermann, K., Dreizler, S., \& Hessman, F. V. 2013, A\&A, 555, 133
%
\reference{ } Bradt, H. 2004, Astronomy Methods, Cambridge Univ. Press, Cambridge, p.170
%
\reference{ } Ceraski, A. V. 1903, Astron. Nach., 161, 159
%
\reference{ } Cumming, A. 2004, MNRAS, 354, 1165
%
\reference{ } Cumming, A., Marcy, G. W., \& Butler, R. P. 1999, ApJ, 526, 890
%
\reference{ } Dormand, J. R., El-Mikkawy, M. E. A., \& Prince, P. J. 1987,
IMA J. Numer. Anal., 7, 423
%
\reference{ } Duffett-Smith, P., \& Zwart, J. 2011, Practical Astronomy with your Calculator or Spreadsheet,
Cambridge Univ. Press, Cambridge, p.31
%
\reference{ } Dugan, R. S. 1915, MNRAS, 75, 702
%
\reference{ } Eastman, J., Siverd, R., \& Gaudi, B. S. 2010, PASP, 122, 935
%
\reference{ } Go\'{a}dziewski, K., Nasiroglu, I.,  S{\l}owikowska, A., Beuermann, K., \& Kanbach, G. 2012, MNRAS, 425, 930
%
\reference{ } Go\'{a}dziewski, K., S{\l}owikowska, A., Dimitrov, D., Krzeszowski, K., \& \.{Z}ejmo, M. 2015, MNRAS, 448, 1118
%
\reference{ } Irwin, J. B. 1952, ApJ, 116, 211
%
\reference{ } Kreiner, J. M., Kim, C.-H., \& Nha, I.-S. 2001, An Atlas of O ¨C C Diagrams
of Eclipsing Binaries (Krakow: Wydawnictwo Naukowe Akademii Pedagogicznej)
%
\reference{ } Kwee, K. K., \& van Woerden, H. 1956, Bull. Astron. Inst. Netherlands, 464, 327
%
\reference{ } Markwardt, C. B. 2009, in Bohlender D. A., Durand D., Dowler P., eds, ASP Conf. Ser., Vol. 411,
Astronomical Data Analysis Software and Systems XVIII. Astron. Soc. Pac., San Francisco, p.
251
%
\reference{ } Marsh, T. R., Parsons, S. G., Bours, M. C. P., Littlefair, S. P., Copperwheat, C. M., Dhillon, V. S., Breedt, E., Caceres, C., \& Schreiber, M. R. 2014, MNRAS, 437, 475
%
\reference{ } Qian, S.-B. 2001, MNRAS, 328, 914
%
\reference{ } Qian, S.-B. 2003, MNRAS, 342, 1260
%
\reference{ } Qian, S.-B. 2008, AJ, 136, 2493
%
\reference{ } Rafert, J. B. 1982, PASP, 94, 485
%
\reference{ } Struve, O. 1947, ApJ, 106, 92
%
\reference{ } Terrell, D. 2006, Inf. Bull. Var. Stars, 5742
%
\reference{ } Yuan, J.-Z., \& Qian, S. B. 2007, ApJ, 669, L93
%
\reference{ } Yuan, J.-Z., \& \c{S}enavc{\i}, H. V. 2014, MNRAS, 439, 878 (Paper I)
%
\reference{ } Zechmeister, M., \& K\"{u}rster, M. 2009, A\&A, 496, 577
%
\reference{ } Zhang, X.-J., Li, H., Li, S.-T., \& Lin, D. N. C. 2014, ApJ, 789, L23
%
\end{references}
\end{document}